\documentclass[aps,prb,reprint,superscriptaddress]{revtex4-1}

\usepackage{subfigure}
\usepackage{epsfig}
\usepackage{float}
\usepackage{multirow}
\usepackage{url}

\usepackage{graphicx}
\usepackage{amssymb, amsmath}
\usepackage[usenames]{color}
\usepackage{dcolumn}
\usepackage{bm}
\usepackage{ textcomp }

\begin{document}

\title{Theory of spin Bott index for quantum spin Hall states in non-periodic systems}

\author{Huaqing Huang}
\affiliation{Department of Materials Science and Engineering, University of Utah, Salt Lake City, Utah 84112, USA}

\author{Feng Liu\footnote{Corresponding author: fliu@eng.utah.edu}}
\affiliation{Department of Materials Science and Engineering, University of Utah, Salt Lake City, Utah 84112, USA}
\affiliation{Collaborative Innovation Center of Quantum Matter, Beijing 100084, China}

\date{\today}

\begin{abstract}
This is a joint publication with the Letter by H. Huang and F. Liu [Phys. Rev. Lett. 121, 126401 (2018)]. In this work, we propose the spin Bott index to identify the quantum spin Hall (QSH) state in both crystalline and non-periodic systems. The applicability of the spin Bott index is confirmed by analyzing the periodic and disorder Kane-Mele models. As an example of non-periodic systems, we systematically investigate the QSH effect in a Penrose-type quasicrystal lattice (QL). We characterized the nontrivial electronic topology of the QL by directly calculating the spin Bott index. In addition, the topological edge states, the localization of wavefunctions and quantized transport signatures are also studied in detail.
\end{abstract}

\maketitle

\section{Introduction}
Since the discovery of $\mathbb{Z}_2$ topological insulators by Kane and Mele,\cite{PhysRevLett.95.146802,PhysRevLett.95.226801} several methods have been proposed to calculate the $\mathbb{Z}_2$ index. In particular, for systems with inversion symmetry, Fu and Kane simplify the calculation of $\mathbb{Z}_2$ by considering the parity of occupied states at time-reversal invariant points in the Brillouin zone.\cite{parity} For general time-reversal-invariant systems, they also derive an efficient formula for the $\mathbb{Z}_2$ index which is expressed as an integral involving the Berry connection and Berry curvature. \cite{PhysRevB.74.195312,JPSJ.76.053702} Soluyanov \textit{et al}. and Yu \textit{et al}. develop an effective method to determine $\mathbb{Z}_2$ index based on the evolution of hybrid Wannier charge centers. \cite{alexey2,yuruiZ2} In addition, a spin Chern number was also suggested to characterize the $\mathbb{Z}_2$ topological order. \cite{PhysRevLett.97.036808,PhysRevB.75.121403,PhysRevB.80.125327} However, all these methods are only applicable for periodic systems. For non-periodic systems, an effective numerical determination of $\mathbb{Z}_2$ index is more challenging.
So far, several numerical methods have been proposed for non-periodic systems. For example, Kane and Mele propose to determine the $\mathbb{Z}_2$ index by a certain Pfaffian with twisted boundary condition. \cite{PhysRevLett.95.146802,PhysRevB.76.165307,PhysRevB.85.205136,PhysRevB.83.235115} Another method is based on scattering matrix theory, in which the $\mathbb{Z}_2$ index is defined by the scattering matrices at the Fermi level without the knowledge of the full spectrum. \cite{PhysRevB.83.155429,PhysRevB.85.165409,PhysRevB.89.155311} Ringel and Kraus provided an algorithm for determining the $\mathbb{Z}_2$ index, which can be extracted from a local equal time ground-state correlation function. \cite{PhysRevB.83.245115} Loring and Hastings extended the definition of $\mathbb{Z}_2$ index as the Kitaev's $\mathbb{Z}/2$ index 
based on the theory of almost commuting matrices.\cite{loring2011disordered,hastings2011topological, loring2014quantitative} Loring further derives formulas and algorithms for Kitaev's invariants of different classes in the periodic table for topological insulators and superconductors \cite{PhysRevB.78.195125,Kitaev2009periodic, Shinsei2010topological} for finite disordered systems on lattices with boundaries.\cite{loring2015k} Meanwhile, the concept of spin Chern number is also extended to disordered system based on the non-commutative theory of Chern number. \cite{prodan2010non,prodan2011disordered}

In this work, we define an alternative topological invariant, i.e., the spin Bott index, to determine the QSH state in both periodic and non-periodic systems. It is based on previous works on Bott index \cite{bellissard1994noncommutative,hastings2010almost,exel1991invariants} and spin Chern number \cite{PhysRevLett.97.036808, PhysRevB.75.121403,PhysRevB.80.125327,prodan2010non,prodan2011disordered}. The equivalence of the spin Bott index and the $\mathbb{Z}_2$ invariant is checked using the Kane-Mele model. To test the applicability and effectiveness of our numerical algorithm, we further study the topological Anderson insulator state in the disorder Kane-Mele model. As an example, we systematically investigate the QSH effect in a Penrose-type QL. The QSH state is directly determined by calculating the spin Bott index. In addition, we study the topological edge states, the localization of wavefunctions and transport properties which further confirm the nontrivial topological character of the QL.

This paper is organized as follows. In Sec.~\ref{sec_bott}, we derive the general properties of spin Bott index. In Sec.~\ref{sec_model}, we introduce the details of the model that will be used for illustrative calculations of QLs. The numerical results and discussion of QSH effect in QLs are presented in Sec.~\ref{sec_results}, and we end with a summary in Sec.~\ref{sec_conclusion}.

\section{\label{sec_bott}Bott index and spin Bott index}
In this section, we first present the calculation details about the Bott index and the spin Bott index we proposed (Sec.~\ref{method}).
Then we use the Haldane model as an example to illustrate the relationship between the Bott index and the Chern number in Sec.~\ref{haldane}. Next, we use the Kane-Mele model to illustrate the equivalence of the spin Bott index and the $\mathbb{Z}_2$ invariant in Sec.~\ref{kanemele}. Finally, we calculate the spin Bott index of our model in a disordered lattice in Sec.~\ref{random}, demonstrating the applicability of spin Bott index to non-periodic systems.

\subsection{\label{method}Method of calculation}
\begin{figure*}
\includegraphics[width =1\textwidth]{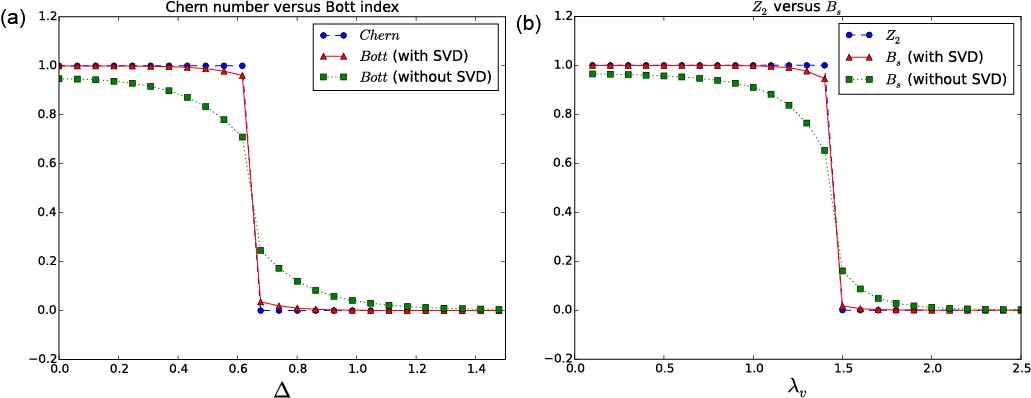}%
\caption{\label{figs_BsB} (a) Topological phase transition in the Haldane model. The parameters are $t=-1$ and $t_2=0.15e^{-i\pi/3}$. (b) Topological phase transition in the Kane-Mele model. The parameters are $t=1,\lambda_{SO}=0.3$ and $\lambda_R=0.25$. The Bott $B$ (spin Bott $B_s$) index is consistent with the Chern number ($\mathbb{Z}_2$ invariant) except around the phase transition point. This is because we use a relatively small supercell to calculate the Bott index. The small divergence would disappear if a larger supercell is used in the calculation of (spin) Bott index. The calculated (spin) Bott index with SVD shows a better performance than the one without SVD.}
\end{figure*}

The method of calculating the Bott index has already been explained in the literature. \cite{loring2011disordered,hastings2011topological, loring2015k,PhysRevLett.118.236402,PhysRevX.6.011016} Firstly, one constructs the projector operator of the occupied states below a given gap
\begin{equation}
P=\sum_i^{N_{occ}}|\psi_i\rangle\langle\psi_i|,
\end{equation}
where $|\psi_i\rangle$ is the wavefunction of the $i$-th state with eigenvalue $\epsilon_i$.
Next, one calculates the projected position operators,
\begin{eqnarray}
U=Pe^{i2\pi X}P,\\
V=Pe^{i2\pi Y}P,
\end{eqnarray}
where $X$ and $Y$ are the rescaled coordinates which are defined in the interval $[0,1)$. The Bott index, which measures the commutativity of the projected position operators, \cite{bellissard1994noncommutative,hastings2010almost,exel1991invariants,katsura2016Z2,*katsura2018noncommutative} is given by,
\begin{equation}
B=\frac{1}{2\pi}\textrm{Im}\{\textrm{tr}[\log(VUV^\dag U^\dag)]\}.
\end{equation}
The Bott index is proved to be equivalent to Chen number. \cite{toniolo2017equivalence} Therefore it serves as a topological invariant to distinguish topological nontrivial from trivial states.

Given the method of calculating the Bott index, now we give a general construction of the spin Bott index. One begins by introducing a projected spin operator
\begin{equation}
P_z=P\hat{s}_zP,
\end{equation}
where $\hat{s}_z=\frac{\hbar}{2}\sigma_z$ is the spin operator ($\sigma_z$ is the Pauli matrix). For a spin conserving model, $\hat{s}_z$ commutes with the Hamiltonian $H$ and $P_z$, the Hamiltonian as well as eigenvectors can be divided into spin-up and spin-down sectors. Thus, the eigenvalues of $P_z$ consist of just two nonzero values $\pm\frac{\hbar}{2}$. For systems without spin conservation (for example, the Kane-Mele model with nonzero Rashba terms which will be discussed later), the $\hat{s}_z$ and $H$ no longer commute. The spectrum of $P_z$ spreads toward zero. However, as long as the spin mixing term is not too strong, the eigenvalues of $P_z$ remain two isolated groups which are separated by zero. Since the rank of $P_z$ is $N_{occ}$, the number of positive eigenvalues equals to the number of negative eigenvalues, which is one half of $N_{occ}$.
The corresponding eigenvalue problem can be denoted as
\begin{equation}
P_z|\pm \phi_i\rangle=S_\pm|\pm\phi_i\rangle.
\end{equation}
In this way one can construct new projector operators,
\begin{equation}
P_\pm=\sum_i^{N_{occ}/2}|\pm\phi_i\rangle\langle\pm\phi_i|,
\end{equation}
which satisfy $P=P_+\oplus P_-$, and projected position operators
\begin{eqnarray}
U_\pm=P_\pm e^{i2\pi X}P_\pm+(I-P_\pm),\label{eqU}\\
V_\pm=P_\pm e^{i2\pi Y}P_\pm+(I-P_\pm),\label{eqV}
\end{eqnarray}
for two spin sectors, respectively.

It is noted that the complementary projectors $Q_\pm=1-P_\pm$ are added into the definition of projected position operators, \cite{toniolo2017topological} which does not change the final results but makes the numerical algorithm more stable. The Bott index measures the commutativity of a pair of almost commuting and almost unitary matrices, which can distinguish the pairs of commuting matrices close to commuting pairs from those are far from commuting pairs. \cite{exel1991invariants,loring2014quantitative,loring2011disordered} Adding the complementary projectors $Q_\pm$ into Eq.~(\ref{eqU}) and Eq.~(\ref{eqV}) makes the position operator close to a unitary matrix, which is useful in numerical calculations. For a better convergency of the numerical algorithm, the product $U_\pm V_\pm U_\pm^\dag V_\pm^\dag$ should be unitary. Therefore, $\det(U_\pm V_\pm U_\pm^\dag V_\pm^\dag)=1$ and the spectrum of $\log(U_\pm V_\pm U_\pm^\dag V_\pm^\dag)$ is purely imaginary, then the Bott index is a real integer. \cite{toniolo2017equivalence} To further increasing the stability of the numerical algorithm, one performs a singular value decomposition (SVD) $M=Z\Sigma W^\dag$ for the projected position operators $U_\pm$ and $V_\pm$, where $Z$ and $W$ are unitary and $\Sigma$ is real and diagonal. Then one can identify the ``unitary part'' $\tilde{M}=ZW^\dag$ as the new projected position operators which are unitary now. Mathematically, the SVD process is equivalent to a scaling transformation which does not change the commutativity of two projected position operators. Therefore, applying SVD does not obscure an actual breakdown of the original formalism, but effectively improves the convergence and stability of the numerical algorithm, as shown later (see Fig.~\ref{figs_BsB}).

The Bott indices for two spin sectors are now given by\cite{bellissard1994noncommutative, hastings2010almost, exel1991invariants, katsura2016Z2, *katsura2018noncommutative}
\begin{equation}
B_\pm=\frac{1}{2\pi}\textrm{Im}\{\textrm{tr}[\log(\tilde{V}_\pm \tilde{U}_\pm \tilde{V}_\pm^\dag \tilde{U}_\pm^\dag)]\}.
\end{equation}
Finally, we define the spin Bott index as the half difference between the Bott indices for the two spin sectors
\begin{equation}
B_s=\frac{1}{2}(B_+-B_-).
\end{equation}
Similar to the spin Chern number, \cite{PhysRevLett.97.036808,PhysRevB.75.121403,PhysRevB.80.125327,prodan2010non,prodan2011disordered} the spin Bott index is a well-defined topological invariant. The spin Bott index is also directly related to the $\mathbb{Z}_2$ topological invariant. Its robustness is due to the existence of two spectral gaps: the insulating gap of the Hamiltonian and the spectral gap of the projected spin operator $P_z$. As long as the two gaps persist, the computational formalism of the spin Bott index can be applied.
The spin Bott index is applicable to quasiperiodic and non-periodic systems, which provides especially a useful tool to determine the electronic topology of those systems without periodicity.

\begin{figure*}
\includegraphics[width =1\textwidth]{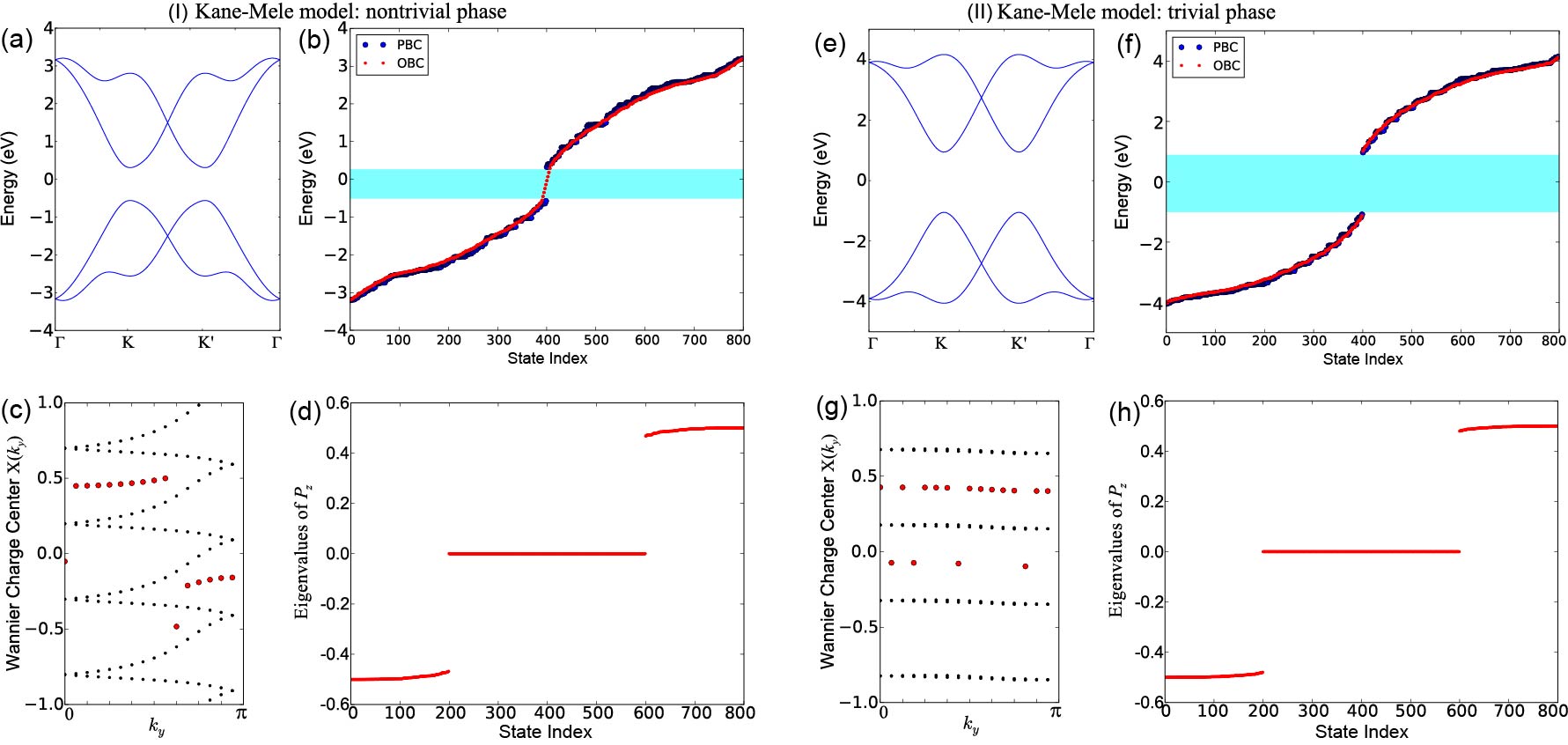}%
\caption{\label{figs_KM} Calculation of Kane-Mele model. The left panel (I) Kane-Mele model with a nontrivial QSH phase. The parameters are $t=1, \lambda_{SO}=0.3, \lambda_R=0.25$ and $\lambda_v=1.0$. The right panel (II) Kane-Mele model with a nontrivial QSH phase. The parameters are $t=1, \lambda_{SO}=0.3, \lambda_R=0.25$ and $\lambda_v=2.5$. In each panel we show (a,e) the band structure of unit cell; (b,f) energy eigenvalues versus state index of supercell sample with periodic boundary condition or open boundary condition; (c,g) evolution of the Wannier charge center $X(k_y)$ along $0 \leqslant k_y \leqslant \pi$ (Red points mark the midpoint of the largest gap); (d,h) eigenvalues of $Ps_zP$ (in units of $\hbar$) versus state index.}
\end{figure*}

\subsection{\label{haldane}Relationship between Bott index and Chern number}
To illustrate the equivalency of the Bott index and the Chern number, we use the Haldane model \cite{PhysRevLett.61.2015,PhysRevB.74.235111} as an example. The Haldane model exhibits a nonzero quantization of the Hall conductance in the absence of an external magnetic field.
The Hamiltonian is written as
\begin{equation}
H=t\sum_{\langle ij\rangle}c_i^\dag c_j+t_2\sum_{\langle\langle ij\rangle\rangle}\nu_{ij}c_i^\dag c_j+\Delta \sum_i\xi_ic_i^\dag c_i.
\end{equation}
The first term is a nearest-neighbor (NN) hopping term on the honeycomb lattice. The second term is a next-NN hopping term, which carries a phase factor. The last term is a staggered sublattice potential (where $\xi_i=\pm1$). The system undergoes a transition from the Chern-insulator to the normal-insulator phase when tuning the staggered potential ($\Delta$) in the last term.

As shown in Fig.~\ref{figs_BsB}(a), with the increasing $\Delta$, the Chern insulator phase with $C=1$ becomes a normal insulator with $C=0$. The calculated Bott index shows similar behavior, except for a small discrepancy around the phase transition point. According to Ref.~\onlinecite{toniolo2017equivalence}, the difference between the Chern number and Bott index is within a correction of the order $O(1/L)$, where $L$ is the system size. Because the energy gap reduces to zero and the correlation length increases dramatically near the phase transition point, it requires a larger sample size to reach a high accuracy of the Bott-index calculation. Therefore, finite size effect induces a small divergence between the Chern number and Bott index. However, one can still easily distinguish topological nontrivial from trivial states as the divergence is quite small. Furthermore, by increasing the sample size, one can get a more accurate Bott index even around the phase transition point. Comparing the Bott index calculated with and without SVD, it is clear that the Bott index calculated without SVD departs from the exact value of the topological invariant, even when the system is away from the phase transition. Moreover, with the increasing sample size, the method with SVD converges faster than the one without SVD, to the exact value of the topological invariant. This indicates that applying the SVD does not destroy the original formalism, but indeed improves the convergence and stability of the numerical algorithm.

\subsection{\label{kanemele}Relationship between spin Bott index and $\mathbb{Z}_2$ invariant}
\begin{figure}
\includegraphics[width =1\columnwidth]{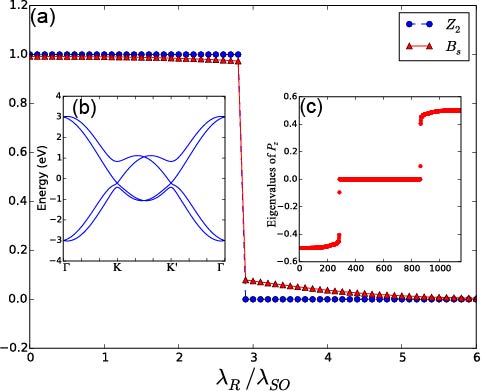}
\caption{\label{figs_Rashba} Topological phase transition in the Kane-Mele model. The parameters are $t=1$, $\lambda_v=0.1t$, $\lambda_{SO}=0.06t$.\cite{PhysRevLett.95.146802} (a) The $\mathbb{Z}_2$ invariant and spin Bott index $B_s$ versus $\lambda_R/\lambda_{SO}$. The inset shows the (b) band structure and (c) eigenvalues of $P_z$ (in units of $\hbar$) upon phase transition with $\lambda_R/\lambda_{SO}=2.86$, where both the band gap of the Hamiltonian and the spectral gap of $P_z$ almost vanish.
}
\end{figure}

To check the relationship between the spin Bott index and $\mathbb{Z}_2$ index, we adopt the Kane-Mele model \cite{PhysRevLett.95.146802,PhysRevLett.95.226801} as an example.
The Kane-Mele Hamiltonian on a graphene lattice is given by
\begin{eqnarray}
H_{\mathrm{KM}}&=&t\sum_{\langle ij\rangle}c_i^\dag c_j+i\lambda_{SO}\sum_{\langle\langle ij\rangle\rangle}\nu_{ij}c_i^\dag s_z c_j\nonumber\\
&+&i\lambda_R\sum_{\langle ij\rangle}c_i^\dag (\mathbf{s}\times\hat{\mathbf{d}}_{ij})_z c_j+\lambda_v\sum_i\xi_ic_i^\dag c_i.
\label{eq_KM}
\end{eqnarray}
The first term is a NN hopping term on the honeycomb lattice, where we have suppressed the spin index on the electron operators. The second term is the mirror symmetric spin-orbit coupling (SOC) term which involves spin dependent second NN hopping. Here $\nu_{ij}=\frac{2}{\sqrt{3}}(\hat{\mathbf{d}}_1\times \hat{\mathbf{d}}_2)$, where $\hat{\mathbf{d}}_1$ and $\hat{\mathbf{d}}_2$ are unit vectors along the two bonds that the electron traverses going from site $j$ to $i$. $s_z$ is a Pauli matrix describing the electron spin. The third term is a NN Rashba term, which explicitly violates the $M_z$ mirror symmetry. This term will arise from a perpendicular electric field or interaction with a substrate. The fourth term is a staggered sublattice potential ($\xi_i=\pm1$), which can be included to describe the transition between a QSH phase and a normal insulator. This term violates the symmetry under twofold rotations in the plane.
By tuning the staggered sublattice potential ($\lambda_v$) in the last term, one can realize the transition between a topologically nontrivial QSH phase and a trivial insulator, as shown in Fig.~\ref{figs_BsB}(b). We calculate the $\mathbb{Z}_2$ invariant by directly tracing the evolution of one-dimensional (1D) hybrid Wannier charge centers (WCCs) \cite{wannier1} during a ``time-reversal pumping" process. \cite{alexey2} By increasing $\lambda_v$, the QSH insulator with $\mathbb{Z}_2=1$ is driven to a trivial insulator with $\mathbb{Z}_2=0$. The calculated spin Bott index $B_s$ is consistent with the $\mathbb{Z}_2$ invariant. The small divergence between the $\mathbb{Z}_2$ invariant and $B_s$ around the phase transition point is induced by the finite size of the sample, which is similar to that between the Chern number and the Bott index.
Also, the spin Bott index calculated with SVD shows a better performance than the one without SVD, similar to the case of the Bott index.

In Fig.~\ref{figs_KM}, we give two specific examples of the Kane-Mele model in different phases. The calculated $\mathbb{Z}_2$ invariant are 1 and 0 for the QSH phase ($\lambda_v=1.0$) and the trivial insulator ($\lambda_v=2.5$), respectively. We calculated the spin Bott index using a $10\times 10\sqrt{3}$ lattice with the same Hamiltonian and under periodic boundary condition (PBC). The spectra of the projected spin operator $P_z=P\hat{s}_zP$ are separated into three groups for both trivial and nontrivial cases as shown in Fig.~\ref{figs_KM}(d) and ~\ref{figs_KM}(h). In the spectrum of $P_z$, $N_{occ}/2$ eigenvalues are positive, $N_{occ}/2$ eigenvalues are negative and the rest are zero. It is noted that the negative (positive) branch of the spectrum is not a straight line at $-\frac{1}{2}$ ($\frac{1}{2}$) but slightly increases with the state index. This is because the nonzero Rashba term in the Kane-Mele model mix the spin up and spin down states breaking down the $\hat{s}_z$ conservation. The calculated spin Bott index are $0.9957\thickapprox 1$ and $2.1581\times 10^{-5}\thickapprox 0$ for the QSH phase and the trivial insulator, respectively. These results indicate that our proposed method is in good agreement with the $\mathbb{Z}_2$ topological invariant.

It is known that the presence of nontrivial $\mathbb{Z}_2$ invariant corresponds to the existence of topological protected boundary states in a system with open boundary condition (OBC). The finite sample is simulated using a $10\times 10\sqrt{3}$ supercell with both PBC and OBC. The sample contains 400 atoms, which is large enough to show the existence of topological edge state. As shown in Fig.~\ref{figs_KM}(b), the edge states are in the gapped region of the PBC calculation which corresponds to the bulk gap.

To check the robust applicability of the spin Bott index, we also studied the evolution of both the $\mathbb{Z}_2$ invariant and the spin Bott index with the increasing Rashba term in Eq.~(\ref{eq_KM}) which violates the spin conservation. It is known that the spin up and down channels are mixed and a phase transition from a QSH to a normal insulator can be induced by increasing the Rashba term.\cite{PhysRevLett.95.146802} As shown in Fig.~\ref{figs_Rashba}, the spin Bott index is consistent with the $\mathbb{Z}_2$ invariant with a tiny error in most cases. The negligible discrepancy occurs only around the phase transition point where $\lambda_R/\lambda_{SO}\approx3$. This is because both the energy gap of the Hamiltonian and the spectral gap of $P_z$ almost vanish around the phase boundary (see insets of Fig~\ref{figs_Rashba}), then the computational formalism of spin Bott index is no longer applicable.
It is also worth noting that the spin Bott index agree well with the $\mathbb{Z}_2$ invariant even around the phase transition induced by $\lambda_v$ as shown in Fig.~\ref{figs_BsB}(b). Because only the energy gap vanishes, the spectral gap of $P_z$ persists around the phase transition point induced by $\lambda_v$. Therefore, the applicability of the spin Bott index is robust against spin-mixing perturbations ($s_z$-nonconserving terms), which is guaranteed by the coexistence of the energy gap and the $P_z$ spectral gap.

\begin{figure}
\includegraphics[width =1\columnwidth]{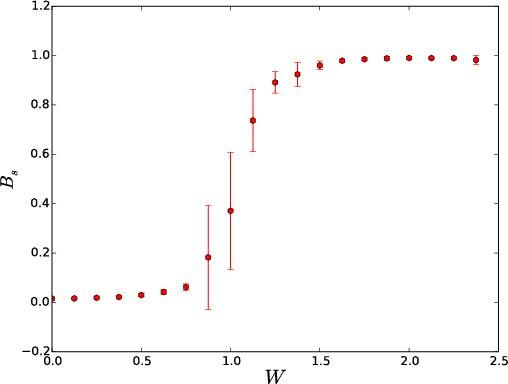}%
\caption{\label{figs_disorder} The spin Bott index as a function of disorder strength $W$ in the disorder Kane-Mele model. The parameters are $t=1$, $\lambda_v=1.65 t$, $\lambda_{SO}=0.3 t$ and $\lambda_R=0$.\cite{orth2016topological} For each $W$, 10 samples with 1600 atoms are used to calculate the spin Bott index. A disorder-induced topological phase transition occurs around $W=1.0t$. This is consistent with the conductance calculation in previous work. \cite{orth2016topological}}
\end{figure}

\subsection{\label{random}Application of spin Bott index in disordered lattices}
To further investigate the applicability and effectiveness of our formulation of the spin Bott index in non-periodic systems, we study the disorder Kane-Mele model, namely, the Eq.~(\ref{eq_KM}) with an additional on-site Anderson disorder term:
\begin{equation}
H_{\mathrm{disorder}}=H_{\mathrm{KM}}+W\sum_i\epsilon_ic_i^\dag c_i,
\end{equation}
where $W$ is the disorder strength and $\epsilon_i$ is a uniformly distributed random variables in $[-1,1)$.\cite{orth2016topological} This model is used to study the topological Anderson insulator phenomena, \cite{PhysRevLett.102.136806, PhysRevLett.103.196805, PhysRevB.84.035110} in which a disorder-induced transition into a phase of quantized conductance occurs.

As shown in Fig.~\ref{figs_disorder}, a disorder-induced topological phase transition occurs around $W=1.0t$, which is consistent with the conductance calculations in previous works.\cite{orth2016topological} The calculation of spin Bott index is performed in samples with 1600 atoms, and 10 samples are simulated for every $W$. A higher accuracy can be achieved by adopting samples with larger size and doing the statistical average for more samples. Our results confirm the topological Anderson insulator phase in the disorder Kane-Mele model, indicating that the spin Bott index is applicable to disordered systems.

\section{\label{sec_model}The model of quasicrystal lattices}
\begin{figure}
\includegraphics[width =1\columnwidth]{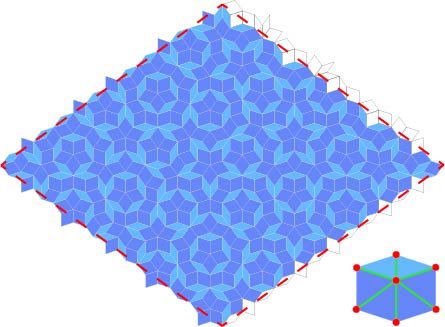}%
\caption{\label{fig1} Penrose tiling containing 521 vertices. The red dashed line defines a unit cell under periodic approximation. The inset shows the atomic orbitals placed on vertices of rhombuses and the first three NN hopping between them.}
\end{figure}

Next, we study the QSH state in QLs.\footnote{\label{fn}See the joint publication: Huaqing Huang and Feng Liu, Phys. Rev. Lett. 121, 126401 (2018).} Although a few models of Chern insulators have been studied in QLs, \cite{PhysRevX.6.011016,PhysRevB.91.085125,PhysRevLett.116.257002} the QSH state in QLs is rarely investigated. To construct the QL, we arrange the lattice sites according to the Penrose tiling consisting of two types of rhombuses (thin and fat).\cite{penrose1974role} Because in the Penrose tiling the thin and fat rhombuses tile the 2D plane completely in an aperiodic way, the QL possesses long-range positional order but lacks the periodicity. Therefore, one cannot use the unit-cell simulation, Brillouin zone, and Bloch theorem as for the crystal calculations. However, it is still possible to generate a series of periodic lattice with growing number of atoms that approximate the infinite quasicrystalline lattice in a systematic way, which is called the quasicrystal tiling approximants. \cite{tsunetsugu1986eigenstates, entin1988penrose, PhysRevB.43.8879, PhysRevX.6.011016} In our model, atomic orbitals are located in the vertices of the Penrose tiling to form a QL, as shown in Fig.~\ref{fig1}. We only consider the first three NN hoppings which are the short diagonal of a thin rhombus, the edge of a rhombus and the short diagonal of a fat rhombus [see inset of Fig.~\ref{fig1}]. If we take the edge length of rhombuses as the unit of length, then the proportion of the three distances are: $r_0:r_1:r_2=2\cos\frac{2\pi}{5}: 1: 2\sin\frac{\pi}{5}$, respectively.

We consider a general atomic-basis model for QLs with three orbitals ($s,p_x,p_y$) per site. The Hamiltonian is given by
\begin{eqnarray}
H_{\mathrm{QL}}&=&\sum_{i\alpha}\epsilon_\alpha c_{i\alpha}^\dag c_{i\alpha}+\sum_{<i\alpha,j\beta>}t_{i\alpha,j\beta}c_{i\alpha}^\dag c_{j\beta}\nonumber\\
 &+&i\lambda\sum_i(c_{ip_y}^\dag \sigma_z c_{ip_x}-c_{ip_x}^\dag \sigma_z c_{ip_y}),
\label{eq1}
\end{eqnarray}
where $c_{i\alpha}^\dag=(c_{i\alpha\uparrow}^\dag,c_{i\alpha\downarrow}^\dag)$ and $c_{i\alpha}=(c_{i\alpha\uparrow},c_{i\alpha\downarrow})^T$ are electron creation and annihilation operators on the $\alpha(=s,p_x,p_y)$ orbital at the $i$-th site. $\epsilon_\alpha$ is the on-site energy of the $\alpha$ orbital. The second term is the hopping term where $t_{i\alpha,j\beta}=t_{\alpha,\beta}(\mathbf{d}_{ij})$ is the hopping integral which depends on the orbital type ($\alpha, \beta=s, p_x, py$) and the vector $\mathbf{d}_{ij}$ between sites $i$ and $j$. $\lambda$ is the SOC strength and $\sigma_z$ is the Pauli matrix.
In our model, the hopping integral $t_{i\alpha,j\beta}=t_{\alpha\beta}(\textbf{d}_{ij})$ is given by the Slater-Koster parametrization
\begin{equation}
t_{\alpha,\beta}(\textbf{d}_{ij})=\mathrm{SK}[V_{\alpha\beta}(d_{ij}),\hat{\textbf{d}}_{ij}],
\end{equation}
where $\hat{\textbf{d}}_{ij}=(l,m)$ is the unit direction vector. In particular, the formula of $\mathrm{SK}[\cdot]$ for the $s,p_x,p_y$ orbitals in our model is written as
\begin{eqnarray}
t_{ss}&=&V_{ss\sigma},\\
t_{sp_x}&=&lV_{sp\sigma},\\
t_{sp_y}&=&mV_{sp\sigma},\\
t_{p_xp_x}&=&l^2V_{pp\sigma}+(1-l^2)V_{pp\pi},\\
t_{p_yp_y}&=&m^2V_{pp\sigma}+(1-m^2)V_{pp\pi},\\
t_{p_xp_y}&=&lm(V_{pp\sigma}-V_{pp\pi}),
\end{eqnarray}
where $V_{\alpha\beta\gamma}=V_{\alpha\beta\gamma}(d_{ij})$ is the $\gamma (=\sigma,\pi)$ bonding parameter between $\alpha (=s,p_x,p_y)$ orbital at the \textit{i}-th site and $\beta (=s,p_x,p_y)$ at \textit{j}-th site. The distance dependence of the bonding parameters $V_{\alpha\beta\gamma}$ is captured approximately by the Harrison relation: \cite{harrison2012electronic}
\begin{equation}
V_{\alpha\beta\gamma}(d_{ij})=V_{\alpha\beta\gamma,0}\frac{d_0^2}{d_{ij}^2}.
\end{equation}
where $d_0$ is a scaling factor to uniformly tune the bonding strengths. Since only the band inversion between \textit{s} and \textit{p} states of different parities is important for the realization of topological states, we focus mainly on 2/3 filling of electron states hereafter, unless otherwise specified.

\section{\label{sec_results}Results and discussion}
In this section, we present the QSH state in QLs. We firstly calculate the spin Bott index to identify the nontrivial topological nature of the QSH state in Sec.~\ref{sec_Bs}. Then we show the electronic properties of QSH state in QLs including the topological edge states and typical bulk wavefunction distributions in Sec.~\ref{sec_edge} and Sec.~\ref{sec_bulk}. Next, we study the phase transition of the QL and the size effect of quasicrystal approximants in Sec.~\ref{sec_size}. The localization of the wavefunction is illustrated by the participation ratio in Sec.~\ref{sec_part}. Finally, we perform transport simulation of the QL based on non-equilibrium Green's function (NEGF) method in Sec.~\ref{sec_transport}.

\begin{figure}
\includegraphics[width =1\columnwidth]{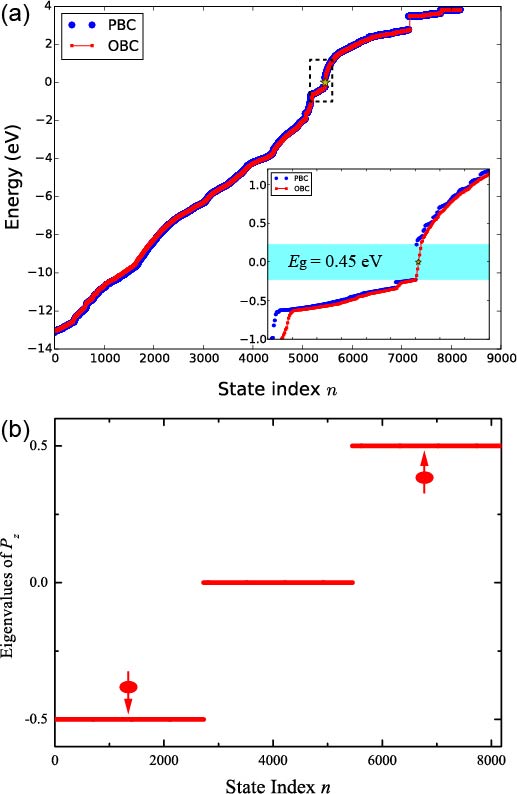}
\caption{\label{figs_spectrum} Calculation of a QL with 1364 atoms. The parameters used here are $\epsilon_s=1.8, \epsilon_{p}=-6.5, \lambda=0.8, V_{ss\sigma}=-0.4, V_{sp\sigma}=0.9,V_{pp\sigma}=1.8$ and $V_{pp\pi}=0.05$ eV. (a) Energy eigenvalues $E_n$ versus the state index $n$. The inset shows the spectrum around the Fermi level. The system with periodic boundary condition (PBC) shows a gap of $E_g=0.45$ eV, while that with open boundary condition (OBC) shows mid-gap states.
(b) Eigenvalues of $Ps_zP$ (in units of $\hbar$) versus state index for the QL.
}
\end{figure}

\subsection{\label{sec_Bs}Spin Bott index of the QL}
Figure~\ref{figs_spectrum}(a) shows the energy eigenvalues of the QL with PBC and OBC, respectively. In the presence of PBC, the system clearly shows an energy gap; i.e, it is an insulator. For the OBC system, there is a set of energy eigenvalues in the mid-gap region, implying the nontrivial electronic topology of the system. To determine the topological nature, we calculated the spin Bott index of the QL with 1364 atoms. The spectrum of the projected spin operator $P_z$ is displayed in Fig.~\ref{figs_spectrum}(b). The eigenvalues are $-\frac{\hbar}{2}$, 0, and $\frac{\hbar}{2}$, respectively. The calculated spin Bott index is $B_s=0.9974 \approx 1$, confirming that the system is a QSH insulator. The nontrivial electronic topology of the bulk spectrum of the QL also indicates the existence of topological edge states on the boundary of finite QL samples. \footnotemark[\value{footnote}]

\begin{figure}
\includegraphics[width =0.9\columnwidth]{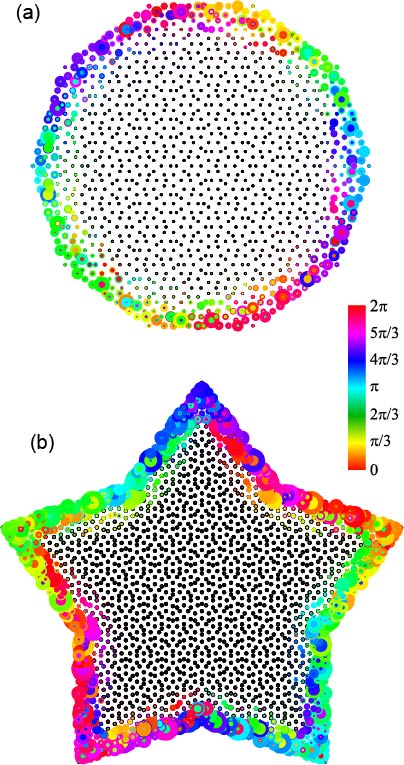}
\caption{\label{figs_circle} Topological edge states on (a) decagon-shape and (b) pentagram-shape QLs with 1211 and 1961 atoms, respectively. The computational parameters are the same as those in Fig.~\ref{figs_spectrum}. The size and the color of the blob indicate the norm $|\rho(\mathbf{r})|^2$ and phase $\phi(\mathbf{r})$ of the wavefunction $|\psi(\mathbf{r})\rangle=\rho(\mathbf{r})e^{i\phi(\mathbf{r})}$, respectively.}
\end{figure}

\begin{figure*}
\includegraphics[width =1\textwidth]{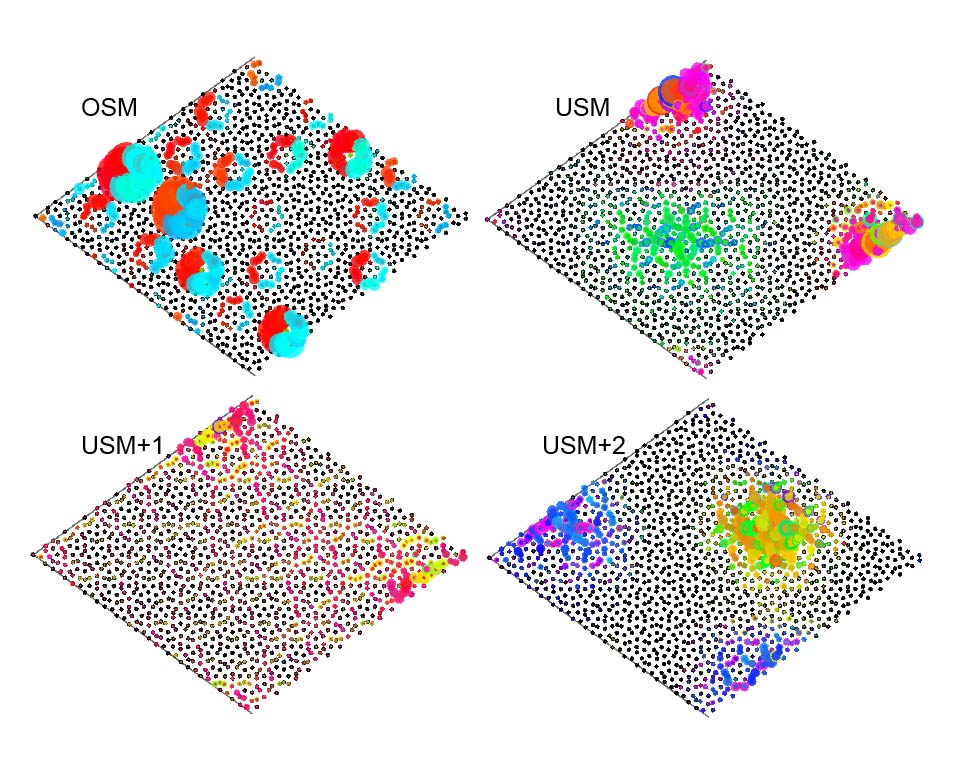}
\caption{\label{figs_QSHpbc} The wavefunction distribution of the occupied state maximum (OSM), unoccupied state minimum (USM) and states above USM of the QSH insulator phase of the QL with periodic boundary condition. The computational parameters are the same as those in Fig.~\ref{figs_spectrum}.}
\end{figure*}

\subsection{\label{sec_edge}Topological edge states at different boundaries}
Due to the bulk-edge correspondence, topological edge states are expected to appear on the boundary of QSH insulators. Indeed, it is found that the mid-gap states are topological edge states which are localized on the perimeter of finite QLs with OBC.\footnotemark[\value{footnote}]
To verify the robustness of the topological edge states in QLs, we studied the real-space distributions of wavefunctions for the mid-gap states in QL samples with decagon and pentagram shapes, respectively. As shown in Fig.~\ref{figs_circle}, the topological edge states are localized at the boundary of these systems regardless of the detailed shapes of the QL samples, indicating that the topological edge states are robust.

\begin{figure*}
\includegraphics[width =0.9\textwidth]{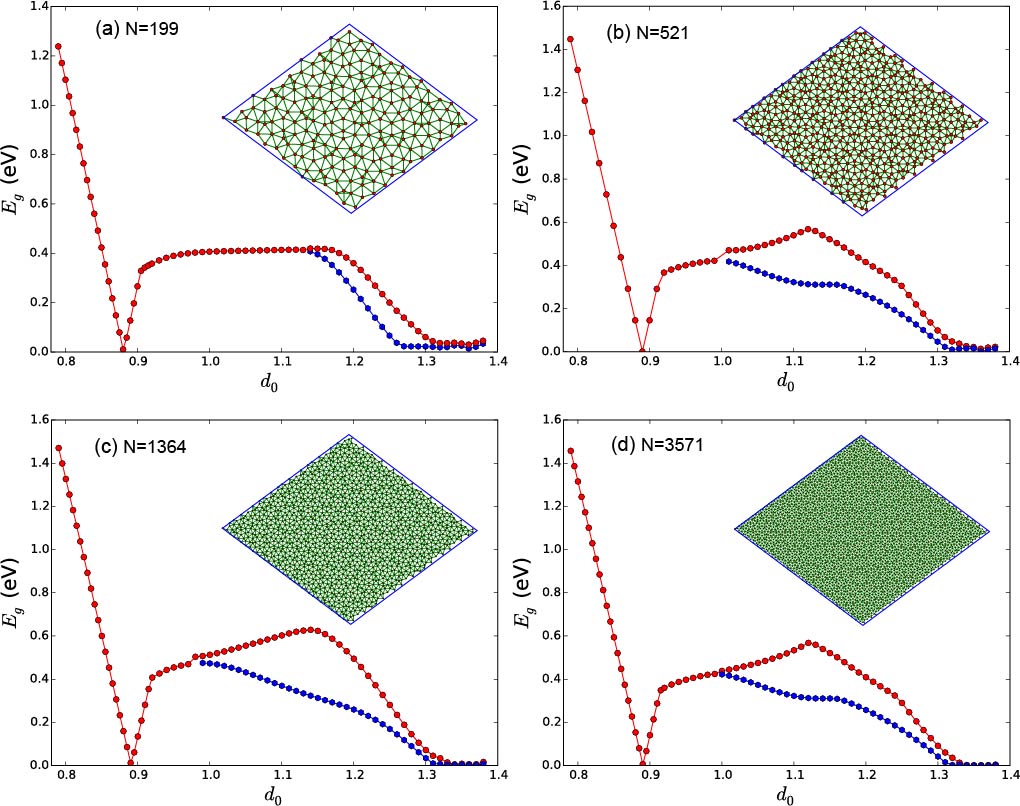}
\caption{\label{figs_gen6789}  Energy gap $E_g$ as a function of interaction strength scale $d_0$ calculated in quasicrystal approximant containing (a) 199, (b) 512, (c) 1364, and (d) 3571 atoms. The phase transitions among NI, QSH state and metal occurs in all approximants with different sizes. The dark blue line in the QSH region represents the defect mode in the energy gap.}
\end{figure*}

\subsection{\label{sec_bulk}Typical wavefunctions of the QL}
Contrary to the extended topological edge state, the typical bulk states of QLs show localized or critical characteristics of quasicrystals. To verify these particular bulk behaviors of QLs, we calculated typical electronic states of quasicrystal approximants with PBC. Figure~\ref{figs_QSHpbc} shows the wavefunctions around the energy gap of the QL in the QSH phase.

In the QL we studied, the basic building blocks of wavefunctions are dimer or trimer states located on sites with the shortest distance ($r_0$, the shortest diagonal of a thin rhombus in the Penrose tiling) \cite{PhysRevB.43.1378}. Some building blocks form confined states or strictly localized states which are referred to as Kohmoto-Sutherland ring state \cite{PhysRevB.34.3849} or Ruby state \cite{PhysRevB.39.9904} [see Fig.~\ref{figs_QSHpbc}(a)]. Besides, several other states called string states, which are self-similar and fractal, can also be formed in Penrose QLs. \cite{PhysRevB.37.2797,PhysRevB.38.1621, PhysRevB.51.15827} Due to the Conway's theorem of Penrose tiling, \cite{penrose1974role} i.e., a finite-size pattern of diameter $d$ is never more than $2d$ away from an exactly identical pattern, these states repeat regularly in the whole QL, which eventually compose the critical wavefunction that is neither extended nor localized. Many unique properties of these critical states, such as self-similarity, power-law decay and gap labeling, have been studied in great detail in the past. \cite{PhysRevB.34.3904,PhysRevLett.56.2740,PhysRevLett.55.2915,PhysRevB.36.6361,PhysRevB.36.7342,PhysRevB.38.5981,PhysRevB.43.8879,PhysRevB.66.094202, PhysRevB.74.054305} These singular electronic properties of critical wavefunctions that originate from the QL structure suggest a poor conductive behavior in the electronic transport properties of quasicrystals. \cite{PhysRevB.35.1456}

\subsection{\label{sec_size}Size effect of quasicrystal approximants}
In general, the spectrum and localization of electronic states of QLs can be strongly affected by the interaction strength between atomic sites. We found that a phase transition among normal insulator (NI), QSH insulator and weak metal (WM) can be realized by increasing the bonding strength uniformly. \footnotemark[\value{footnote}] We further checked the size effect of periodic approximation by calculating the phase diagram of a sequence of quasicrystal approximants with increasing cell size. \cite{tsunetsugu1986eigenstates,entin1988penrose,PhysRevB.43.8879} As shown in Fig.~\ref{figs_gen6789}, similar phase transition among NI, QSH state and WM phase occurs in all quasicrystal approximants. This implies that the topological phase transition as well as the QSH effect should appear in QLs in the thermodynamic limit of infinite lattice size.

\begin{figure*}
\includegraphics[width =0.9\textwidth]{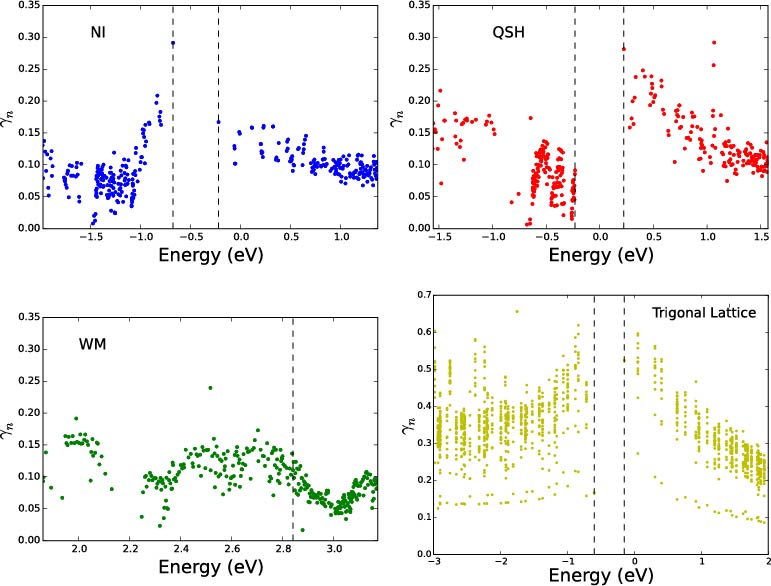}
\caption{\label{figs_Pn} Participation ratio $\gamma_n$ of QLs at (a) normal insulator ($d_0=0.86$),(b) QSH insulator ($d_0=0.95$) and (c) weak metal phases ($d_0=1.34$). (d) Participation ratio $\gamma_n$ of a periodic trigonal lattice.}
\end{figure*}

\subsection{\label{sec_part}Participation ratio}
The conductive behaviors of QLs depend on the localization of wavefunction around the Fermi level. In order to quantify the degree of localization of wavefunctions, we calculate the participation ratio defined by:
\begin{equation}
\gamma_n=\frac{(\sum_i^N|\psi_n^i|^2)^2}{N\sum_i^N|\psi_n^i|^4},
\end{equation}
where the wavefunction $|\psi_n\rangle$ is expended as $|\psi_n\rangle=\sum_i^N \psi_n^i|i\rangle$ on atomic orbital basis $\{|i\rangle\}$. The participation ratio takes the value $1/N$ if the wavefunction is localized in a single orbital and unity if the wavefunction is extended uniformly over the whole system. It is verified that this measure is correlated with the energy transport in the system. \cite{mitchell2018amorphous} We calculated the participation ratio $\gamma_n$ for all three phases of the QL. As shown in Fig.~\ref{figs_Pn}(a)-\ref{figs_Pn}(c), wavefunctions of different phases show relatively small values (less than 0.25 mostly). These are much smaller than those of extended states in periodic crystals [Fig.~\ref{figs_Pn}(d)], indicating that the wavefunctions of QLs are relatively localized. This is consistent with the fact that there are many critical states with power-law decay. The low participation ratio $\gamma_n$ of the gapless phase also suggests a weak metallic behavior in electronic transport, as discussed later.

Although most wavefunctions in QLs are not extended, the localization of states in QLs is different from the situation of Anderson localization induced by disorder. In the Anderson model, the strongest localized states appear at the edges of the mobility gap, while the least localized states with relatively high $\gamma_n$ occur near the gap in the spectrum of QLs [Fig.~\ref{figs_Pn}(a)]. \cite{de2013electronic,odagaki1986properties,schreiber1985numerical} This suggests that the localization behavior of the cases we study here is different from the Anderson localization, and it is probably an effect of the local topology of the Penrose QL. \cite{PhysRevB.51.15827,PhysRevB.34.5208}

\begin{figure}
\includegraphics[width =1\columnwidth]{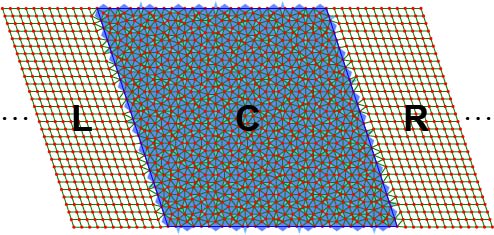}
\caption{\label{figs_transport} Schematic illustration of transport simulation setup using the NEGF method. The central part is a finite QL, the left and right parts are two semi-infinite electric leads.}
\end{figure}

\begin{figure*}
\includegraphics[width =0.8\textwidth]{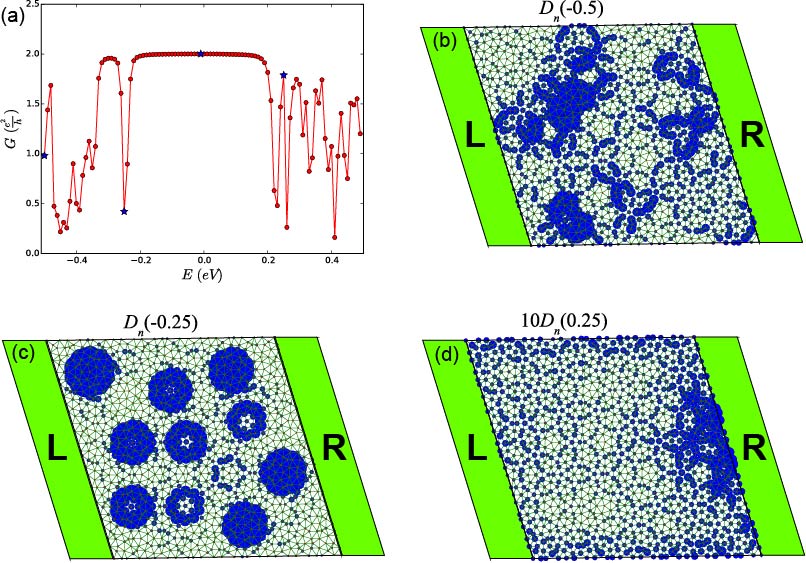}
\caption{\label{figs_QSHtransport} (a) Two-terminal conductance $G$ as a function of the Fermi energy $E$ of quasicrystal in the QSH insulator phase. The conductance $G$ shows a quantized plateau in the energy gap of QSH insulator shown in Fig.2(a) of the main text. (b) Local density of state $D_n(E)$ at $E =-0.5,-0.25$ and $0.25$ eV [marked as stars in (a)] for the central quasicrystal in the transport simulation. The size of blue dot represents the relative value of local density of state. For a better visualization, the local density of states in (d) is rescaled as $10\times D_n(0.25)$.}
\end{figure*}

\begin{figure*}
\includegraphics[width =0.8\textwidth]{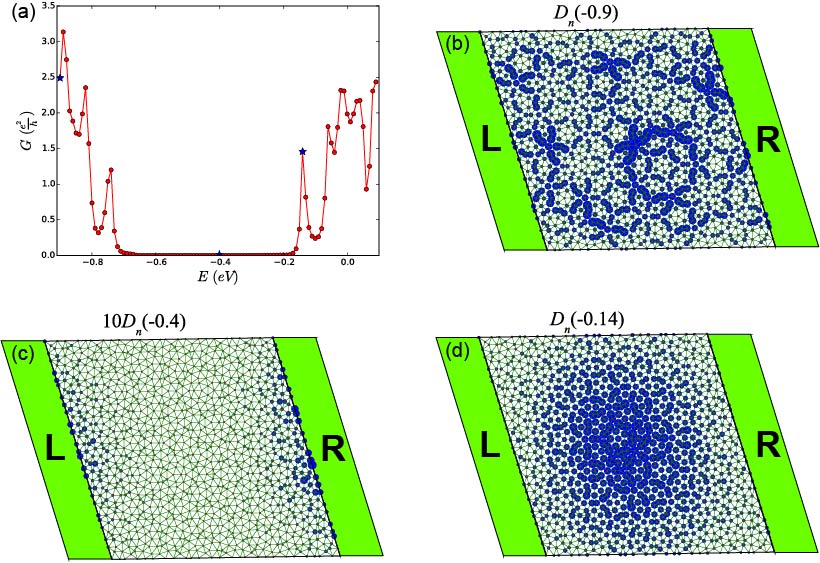}%
\caption{\label{figs_NItransport} (a) Two-terminal conductance $G$ as a function of the Fermi energy $E$ of quasicrystal in the normal insulator phase. The zero-conductance region of $G$ is consistent with the energy gap of normal insulator shown in Fig.3(c) of the main text. (b) Local density of state $D_n(E)$ at $E =-0.9,-0.41$ and $-0.15$ eV [marked as stars in (a)] for the central quasicrystal in the transport simulation. The size of blue dot represents the relative value of local density of state. For a better visualization, the local density of states in (c) is rescaled as $10\times D_n(-0.41)$.}
\end{figure*}

\begin{figure*}
\includegraphics[width =0.8\textwidth]{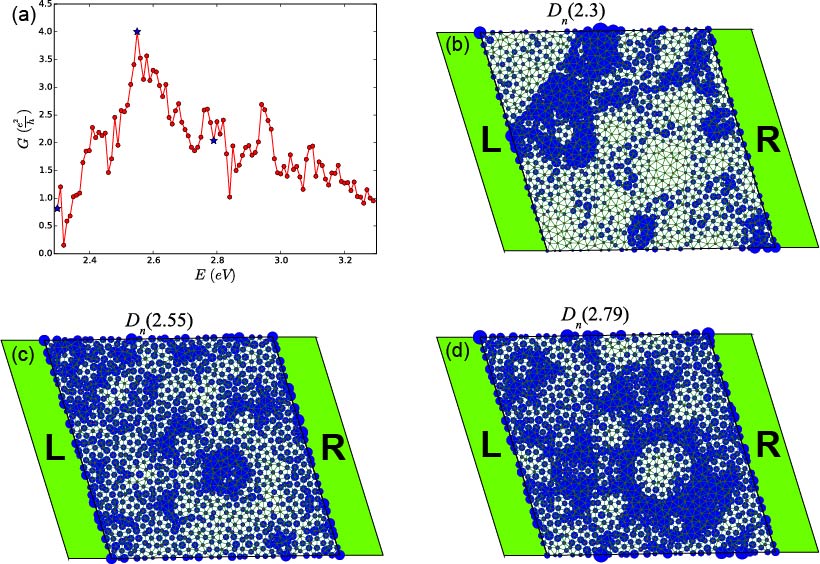}%
\caption{\label{figs_WMtransport} (a) Two-terminal conductance $G$ as a function of the Fermi energy $E$ of quasicrystal in the weak metal phase. The conductance $G$ is much smaller than that of periodic crystal with the same structure of the leads. (b) Local density of state $D_n(E)$ at $E =2.3,2.55$ and $2.79$ eV [marked as stars in (a)] for the central quasicrystal in the transport simulation. The size of blue dot represents the relative value of local density of state.}
\end{figure*}

\subsection{\label{sec_transport}Transport simulation based on NEGF}
Although the above analysis of electronic properties already confirmed the QSH state in the QL, to obtain more details of experimentally measurable quantities, we investigate the transport properties of the QL in different phases. TO do so, we adopt the NEGF method which will be described in Sec.~\ref{negf}. The simulation results are presented in Sec.~\ref{conductance}.

\subsubsection{\label{negf}NEGF formulation}
To investigate the transport properties of the QSH state in QLs, thus demonstrating the metallic edge states, we calculate the two-terminal conductance of the sample using the NEGF method. In the limit of low temperature one may ignore the inelastic backscattering processes, and describe the ballistic transport of the edge states within the Landauer-B\"{u}ttiker \cite{PhysRevB.38.9375} framework. This method is widely used in modeling electron transport through nano-scale devices. In the framework of NEGF, electron transport is treated as a 1D coherent scattering process in the ``scattering region'' for electrons coming in from the electrodes.

In the following, we briefly review the method that we use to simulate the transport process of a finite quasicrystal sample coupled to two semi-infinite periodic leads. In the transport process, the electric current is obtained using the Landauer-B\"{u}ttiker formula, \cite{PhysRevB.38.9375}
\begin{equation}
I=\frac{2e}{h}\int dE T(E)[f_L(E)-f_R(E)],
\label{eq8}
\end{equation}
where $f_{L,R}(E)$ is the Fermi-Dirac distribution for the left or right electrodes. $T(E)$ is the transmission coefficient at energy $E$. It can be calculated from Green's function $G(E)$ of the system.
For the transport system with two semi-infinite leads, the Hamiltonian is given by,
\begin{equation}
H=\left(
\begin{array}{ccccc}
\ddots    & V_L         &             &             &      \\
V_L^\dag  & H_L         & V_{LC}      &             &      \\
          & V_{LC}^\dag & H_C         & V_{CR}      &      \\
          &             & V_{CR}^\dag & H_R         & V_R  \\
          &             &             & V_R^\dag    & \ddots\\
\end{array}
\right).
\end{equation}
The Hamiltonian is composed of left and right semi-infinite leads and the central part.
In principle, the Green's function of the Hamiltonian can be obtained by solving the following equation:
\begin{equation}
(E - H) G(E) = I,
\end{equation}
where $I$ is the identity matrix. However, due to the infinite dimension of the Hamiltonian, it is hard to calculate $G(E)$ directly. As we are only interested in the central part of the transport system, it is more convenient to calculate the Green's function only for the central part and include the effect of leads through self energies $\sum_L$ and $\sum_R$.
Thus, the Green's function of the central part becomes
\begin{equation}
G_C(E)=\left[E-H_C-\Sigma_L-\Sigma_R\right]^{-1},
\end{equation}
where the self energies are
\begin{equation}
\Sigma_{\alpha}=V_{C\alpha}g_{\alpha\alpha}V_{\alpha C}\quad (\alpha=L,R).
\end{equation}
If the lead can be broken down into a semi-infinite stack of principal layers, namely, the lead is within the principal layer approximation, the surface Green's function $g_{\alpha\alpha}$ can be written as \cite{huanghqInterface}
\begin{equation}
g_{\alpha\alpha}=(\epsilon-H_\alpha-V_\alpha \mathcal{T})^{-1},
\end{equation}
where $\epsilon= E + i\eta$ with $\eta$ being arbitrarily small, and the transfer matrix $\mathcal{T}$ is given by
\begin{equation}
\mathcal{T}=(\epsilon-H_\alpha-V_\alpha \mathcal{T})^{-1}V_\alpha^\dag,
\end{equation}
which can be computed self-consistently via an efficient iterative scheme proposed by L\'{o}pez Sancho \textit{et al.} \cite{lopez,lopez2}.

Having the Green's function $G_C(E)$, we now can calculate the transmission function
\begin{equation}
T(E)=\mathrm{Tr}[G_C(E)\Gamma_R(E)G_C(E)\Gamma_L(E)],
\end{equation}
where $\Gamma_{L}$ and $\Gamma_{R}$ are the so-called coupling matrices which are related to the self-energies,
\begin{equation}
\Gamma_\alpha=i(\Sigma_\alpha-\Sigma_\alpha^\dag)\quad(\alpha=L,R).
\end{equation}
Once the Transmission function $T(E)$ is known, the electric current can be easily obtained through Eq.(\ref{eq8}). Moreover, the local density of states of the central part is given straightforwardly with the expression
\begin{equation}
D_n(E)=-\frac{1}{\pi}\mathrm{Im}[\mathrm{Tr}G_C^n(E+i\eta)],
\end{equation}
where $n$ is the index of local atomic sites. The local density of states provide detailed information of the transport process, such as the distribution of transport channels in real space.

\subsubsection{\label{conductance}Transport properties of QLs}
As shown in Fig.~\ref{figs_QSHtransport}(a), the two-terminal conductance shows a quantized plateau at $G=2e^2/h$ when the Fermi energy lies inside the energy gap of the QSH state in the QL, which resembles that of the QSH state in graphene lattice as predicted by Kane and Mele. \cite{PhysRevLett.95.226801} The mid-gap transport channels are formed by topological edge states which are localized at two edges of the central quasicrystal part in the transport simulation [see Fig.2(d) in Ref.~\footnotemark[\value{footnote}]]. However, the transport channels in the bulk state continuum regions are mainly formed by critical states. The local density of states shows a pattern mainly composed of critical wavefunctions with peculiar distributions.

For the normal insulator phase, the electric transport channel in the valence state continuum (e.g. $E=-0.9$ eV) is also formed by the critical states which exhibit self-similarity and critical behaviors, as shown in Fig.~\ref{figs_NItransport}(b).

For the gapless phase, the system shows a metallic behavior in electronic transport (see Fig.~\ref{figs_WMtransport}). However, the calculated conductance is about an order of magnitude smaller than that of periodic crystals with the same structure of the left and right leads (not shown there). This indicates that the gapless phase is a weak metal [see Fig. 3(b) of the main text]. The weak metallic behavior in the transport also agrees with the low participation ratio $\gamma_n$  shown in Fig.~\ref{figs_Pn}(c).

\section{\label{sec_conclusion}Summary}
In summary, we have presented the theory of spin Bott index which is a useful tool to determine the QSH state. The method is stable, has good convergence and can be applicable to both periodic and non-periodic systems, which enable highly efficient numerical calculations. Taking the Kane-Mele model as an example, we show that the spin Bott index is robust against spin-mixing perturbations, which is guaranteed by the coexistence of energy gap and the spectral gap of $P_z$. In order to check the applicability and effectiveness of our algorithm, we have demonstrated the topological Anderson insulator phase in the disorder Kane-Mele model. Furthermore, we investigated the QSH state in a Penrose-type QL and identified its nontrivial topology by obtaining the spin Bott index, and the associated topological edge state and quantized conductance.

\begin{acknowledgments}
This work was supported by DOE-BES (Grant No. DE-FG02-04ER46148). The calculations were done on the CHPC at the University of Utah and DOE-NERSC.
\end{acknowledgments}

\providecommand{\noopsort}[1]{}\providecommand{\singleletter}[1]{#1}%

\end{document}